\documentclass[10pt,conference]{IEEEtran}

\newif\ifDEBUG
\DEBUGfalse

\usepackage[T1]{fontenc}
\usepackage[utf8]{inputenc}
\usepackage{amsmath,amssymb}
\usepackage{booktabs}
\usepackage{graphicx}
\usepackage{xspace}
\usepackage{url}
\usepackage[hidelinks]{hyperref}
\usepackage{enumitem}
\usepackage{xcolor}
\usepackage[most]{tcolorbox}
\usepackage{booktabs}
\usepackage{tabularx}
\usepackage{threeparttable}
\usepackage{array}
\usepackage{multirow}
\let\comment\relax

\ifDEBUG
\usepackage[draft]{commenting}
\else
\usepackage[final]{commenting}
\fi
\usepackage{cleveref}
\declareauthor{Oreofe}{Oreofe}{orange}
\declareauthor{Paschal}{Paschal}{green}
\declareauthor{JD}{JD}{blue}
\declareauthor{Kelechi}{Kelechi}{pink}

\newcommand{\JD}[1]{\comment[JD]{#1}}
\newcommand{\PA}[1]{\comment[Paschal]{#1}}

\newcommand{\toolname}{AVP\xspace}

\newcommand{\myparagraph}[1]{\paragraph{#1}}
\renewcommand{\myparagraph}[1]{\vspace{0.25em} \noindent \hspace{0.085cm}\underline{\textit{#1:}}\xspace}

\usepackage{xspace}
\newcommand{\eg}{\textit{e.g.,}\@\xspace}

\definecolor{RQAccent}{HTML}{3F5F7F}
\definecolor{RQBackground}{HTML}{F4F7FA}

\newtcolorbox{rqanswer}[1]{
enhanced,
breakable,
colback=RQBackground,
colframe=RQAccent,
coltitle=white,
fonttitle=\bfseries\small,
title={#1},
boxrule=0pt,
frame hidden,
borderline west={1.4pt}{0pt}{RQAccent},
sharp corners,
left=6pt,
right=6pt,
top=5pt,
bottom=5pt,
before skip=7pt,
after skip=7pt,
attach boxed title to top left={
xshift=0pt,
yshift=-1.5mm
},
boxed title style={
colback=RQAccent,
colframe=RQAccent,
boxrule=0pt,
sharp corners,
left=5pt,
right=5pt,
top=2pt,
bottom=2pt
}
}

\title{Reproducibility is Not Enough: Artifact Verifiability in Decentralized-Build Package Ecosystems}

\author{
\IEEEauthorblockN{Oreofe Solarin}
\IEEEauthorblockA{
Case Western Reserve University\\
Cleveland, United States\\
ons8@case.edu
}
\and
\IEEEauthorblockN{Kelechi Kalu, James C. Davis, Paschal Amusuo}
\IEEEauthorblockA{
Purdue University\\
West Lafayette, United States\\
\{kalu,davisjam,pamusuo\}@purdue.edu
}
}

\begin{document}
\maketitle

\begin{abstract}
Reproducible and verifiable builds increase trust in distributed software artifacts by enabling independent parties to detect artifacts produced by compromised build or release pipelines. 
However, artifact verification requires more than deterministic builds: a verifier must also recover the source state, build environment, dependencies, and build instructions that produced the artifact. 
Decentralized-build ecosystems make this difficult because artifacts are produced through heterogeneous tools, maintainer-controlled workflows, and fragmented metadata. 
As a result, it remains unclear how often artifacts in these ecosystems can be independently verified.

This paper studies artifact verifiability across four popular decentralized-build package ecosystems. 
We define an independent verifier model that relies only on registry-derivable metadata and an artifact comparison model with tiered equivalence levels. 
We implement these models in an Artifact Verification Pipeline and use it to measure artifact verifiability across the target ecosystems. 
Our results show that, beyond build determinism, verifiability is limited by missing source and build metadata, implicit release transformations, and unconventional build practices. 
Provenance attestations and embedded VCS metadata improve verification, but they do not provide complete rebuild specifications. 
These findings identify concrete metadata gaps and ecosystem-level changes needed to make artifact verification practical at package-registry scale.

\end{abstract}

\begin{IEEEkeywords}
Reproducible builds, artifact verifiability
\end{IEEEkeywords}

\section{Introduction}

Reproducible builds provide an important foundation for trust in released software~\cite{lamb_reproducible_2022,fourne_its_2023,benedetti_empirical_2025}: given the same source and build instructions, an independent party should produce an identical artifact, detecting build-time tampering and confirming that a released artifact corresponds to the reviewed source~\cite{WhyReproducibleBuilds,BestpracticesSoftwaresupplychainOssscbestpracticesmd}. Open-source communities, standards, and government agencies accordingly recommend them as a supply-chain defense~\cite{moore__software_2024,CISANSAODNI2022,sigOpenSourceProject}---especially for ecosystems such as npm and PyPI, which face a growing wave of attacks on compromised build and release pipelines~\cite{snykNodeIpc2026,snykAntV2026,stepsecurityDurabletask2026,snykTanstack2026}.

However, reproducibility alone is insufficient for verifying the integrity of released artifacts~\cite{xiongBuildVerifiabilityJavabased2022,decarnedecarnavaletChallengesImplicationsVerifiable2014}.
An independent verifier must also know the source revision, build environment, and build instructions that produced the artifact in order to reproduce it.
Centralized-build ecosystems such as Debian~\cite{lamb_reproducible_2022} and Nix~\cite{malkaDoesFunctionalPackage2025} make this easier by preserving or publishing artifact-bound build specifications through ecosystem infrastructure.
In contrast, \textit{decentralized-build} ecosystems rely on third-party build infrastructure and diverse release practices, leading to fragmented or missing source and build metadata~\cite{goswami_npm_2020, gao_pyradar_2024}.
Prior work has largely avoided this problem by assuming the relevant metadata is available~\cite{benedetti_empirical_2025}, relying on human augmentation~\cite{google_oss_rebuild}, or treating artifact verification as out of scope~\cite{hassanshahiUnlockingReproducibilityAutomating2025}.
As a result, it remains unclear whether independent artifact verification is feasible in source-oriented package ecosystems.

This paper addresses this gap empirically. We focus on decentralized-build package ecosystems, such as npm and PyPI, because the growing prevalence of software supply-chain attacks highlights the urgency of reproducible builds and verifiability in these ecosystems~\cite{okafor_sok_2022,sonatype_state_2026}.
We define two models for evaluating verifiability: an independent verifier model using only registry-exposed and registry-derived metadata and an artifact comparison model that allows graded levels of artifact equivalence.
We implement these models in \toolname, an Artifact Verification Pipeline that recovers rebuild metadata, reconstructs build environments, rebuilds artifacts, and compares rebuilt outputs against registry artifacts using tiered equivalence levels. 
We use \toolname to measure verification rates, failure causes, and the effects of provenance attestations, artifact characteristics, and development choices across the four ecosystems: Crates.io, npm, PyPI, and RubyGems.

Our results show that artifact verifiability is low and uneven: across ecosystems, roughly 53–87\% of completed rebuilds verify, and recovering the source commit is the dominant obstacle. Provenance attestations improve recovery, but their benefit comes mostly from the development practices of the projects that adopt them rather than the recorded metadata itself; most remaining mismatches trace to unpinned build-tool and dependency versions, CI/CD-generated files, and source modifications, and many are small or metadata-only---byte identity alone is too coarse for source-oriented artifacts. How a package is \emph{built}, its packaging determinism, predicts verifiability far better than what it \emph{is}, with native compilation a structural limit. Finally, against the state-of-practice OSS-Rebuild, \toolname verifies about twice as many artifacts and, unlike curated rebuild platforms, scales to the popular long tail.

In summary, this paper makes three contributions:
\begin{itemize}
\item We define new models for independent artifact verifiability and implement them in \toolname for recovering rebuild metadata, rebuilding and verifying published software artifacts.
\item We empirically study the verifiability of 34{,}005 packages across four source-oriented package ecosystems.
\item We identify the metadata gaps, release practices, and ecosystem-level changes that affect verifiability.
\end{itemize}

\textbf{Significance:}
This paper provides the first cross-ecosystem measurement of independent artifact verifiability. 
Verifiability in decentralized-build ecosystems remains limited by insufficient rebuild metadata. By identifying the metadata gaps and packaging practices that prevent verification, the study provides guidance for registries, standards, and tooling to support verifiability as a practical software supply-chain defense.

\section{Background and Related Concepts}
\label{sec:background}

Here we review
  packages and publishing models,
  associated security risks,
  mitigation by reproducible and verifiable builds,
  and
  the challenges of verifying artifacts in open-source package ecosystems.

\begin{figure}
\centering
\includegraphics[width=0.9\linewidth]{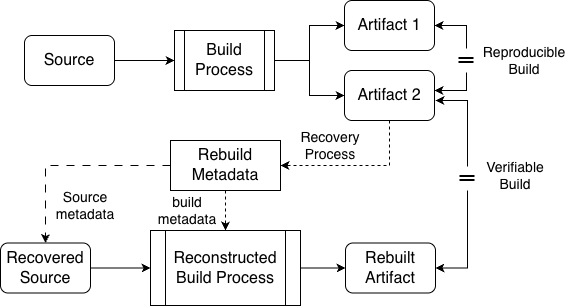}
\caption{
\textbf{Reproducible} builds ensure the same source code produces identical artifacts.
\textbf{Verifiable} builds require an independent party to recover the source revision, build environment, and build instructions needed to reproduce that artifact.
We study the artifact verifiability in decentralized-build ecosystems.}
\label{fig:repro-builds-overview}
\end{figure}





\subsection{Package Ecosystems and Artifacts}
\label{subsec:bg-package-ecosystems}

Software \textit{packages} are reusable software units that provide functionality to downstream applications~\cite{decan_empirical_2019}.
A \emph{package artifact} is the concrete file published for a package version and installed by its users. It contains the package's code, either source directly, or code transformed through compilation, bundling, transpilation, or minification, along with metadata describing the artifact. A \emph{package ecosystem} comprises the packages for a given language or system, the maintainers who publish them, and the tools used to install them~\cite{decan_empirical_2019}.

Ecosystems differ in where artifacts are built. In \emph{centralized-build} ecosystems, typically operating-system distributions, ecosystem-managed infrastructure builds artifacts from submitted source packages---Debian, for example, builds \texttt{.deb} artifacts from signed source packages on its own daemons~\cite{noauthor_debian_nodate}. In \emph{decentralized-build} ecosystems such as npm~\cite{noauthor_npm_nodate} and PyPI~\cite{noauthor_pypi_nodate}, maintainers build artifacts themselves---on their own machines, CI/CD services, or third-party infrastructure---and upload the finished artifact. Because the build runs outside ecosystem-controlled infrastructure, the registry retains no authoritative record of how each artifact was produced~\cite{benedetti_empirical_2025}.

\PA{I removed the paragraph about source-oriented ecosystems since the title no longer uses it. We can also bring back the definition of package ecosystems since it is a key term.}

\subsection{Package Build and Publishing Models and Security Risks}
\label{subsec:bg-build-publish-workflows}

Package ecosystems define the metadata and tooling maintainers use to build and publish artifacts. 
Most require a package manifest in the source directory specifying the name, version, dependencies, and build configuration (\texttt{Cargo.toml} for Crates.io, \texttt{package.json} for npm, \texttt{pyproject.toml} for PyPI), and a \texttt{.gemspec} for RubyGems; in monorepos, the subdirectory holding the manifest identifies the package. 
Ecosystems also provide canonical build and packaging interfaces while permitting third-party build systems and project scripts, such as \texttt{npm pack} for npm or \texttt{python -m build} for PyPI.
To publish, maintainers can either upload pre-built artifacts to registries or publish from CI/CD workflows via trusted publishing~\cite{noauthor_trusted_nodate, crates-trusted-publishing}.

These models expose two security risks~\cite{noauthor_supply_nodate, ohm_backstabbers_2020, duan_towards_2021}. 
In a \emph{build compromise}, an attacker subverts the maintainer's build tools, dependencies, scripts, or CI/CD to inject malicious code during artifact creation; in a \emph{publishing compromise}, an attacker obtains credentials or tokens and uploads a malicious artifact directly. 
Recent npm and PyPI incidents exhibit both~\cite{stepsecurityDurabletask2026, snykTanstack2026, snyk-litellm-2026, noauthor_account_nodate, axiosPostmortem2026, pandyaMaliciousDYdXPackages2026} and in each the distributed artifact diverges from the reviewed source.

\subsection{Reproducible Builds and Artifact Verification}
\label{subsec:bg-repro-builds}
Reproducible and verifiable builds provide independent evidence that a distributed artifact corresponds to its source code. A build is \emph{reproducible} if, given
the same source code, build environment, and build instructions, an independent party can produce a bit-for-bit identical copy of the artifact~\cite{lamb_reproducible_2022, ReproducibleBuildsOrg}.
\emph{Verifiability} adds a recovery requirement: an artifact is verifiable if an independent party, starting only from the published artifact, can recover enough
information about its source and build environment to reproduce an equivalent copy.
The distinction is consequential. Reproducibility \emph{assumes} the source revision, environment, and instructions are known; verifiability requires \emph{recovering} them from the artifact and its metadata. Verifying an artifact therefore rests on two requirements: a build deterministic enough to reproduce, and rebuild information complete enough to drive that reproduction. Because it can detect artifacts altered during build or
publishing, independent verification is recommended as a supply-chain defense by standards bodies~\cite{noauthor_securing_nodate}, government agencies~\cite{CISANSAODNI2022}, researchers~\cite{lamb_reproducible_2022},
and industry~\cite{ReproducibleBuildsOrg}.

The determinism requirement is comparatively well understood: prior work has catalogued the common sources of build non-determinism---timestamps, build paths, filesystem ordering, archive metadata, and embedded randomness~\cite{noauthor_variations_nodate, lamb_reproducible_2022, fourne_its_2023, malkaDoesFunctionalPackage2025}---and the reproducible-builds community has built tooling to remove them, such as \texttt{diffoscope} and \path{SOURCE_DATE_EPOCH}, yielding the high reproducibility rates reported in centrally built ecosystems such as Debian~\cite{lamb_reproducible_2022, DebianReproducibleTests} and Nix~\cite{malkaDoesFunctionalPackage2025}. A complementary line of work relaxes byte-for-byte equivalence, accepting artifacts that preserve relevant structure or behavior~\cite{dietrich_levels_2025}.
The recovery requirement, by contrast, is far less settled, and it is where decentralized package ecosystems diverge from centrally built ones.

\subsection{The Metadata Recovery Gap in Package Ecosystems}
\label{subsec:bg-verifiability-gap}

Even if build non-determinism is controlled, an independent verifier still needs the \emph{rebuild metadata}---the source revision, build environment, and build instructions---that produced the published artifact.
Centrally managed ecosystems can provide this metadata by construction. 
For example, Debian~\cite{lamb_reproducible_2022} and Nix~\cite{malkaDoesFunctionalPackage2025} derive rebuild metadata from their managed build infrastructure and preserve it as part of the artifact or package specification. 
In decentralized-build ecosystems, however, artifacts are often produced outside the registry by maintainer-controlled CI/CD workflows, leaving registries without a complete record of the build inputs and process.

Several efforts attempt to close this metadata recovery gap. 
Maven Reproducible Central rebuilds Maven Central artifacts using contributor-provided build specification files, while recent work explores automated recovery of build metadata and generation of such specifications~\cite{hassanshahiUnlockingReproducibilityAutomating2025,keshani_aroma_2024}. 
These approaches use heuristics and provenance attestations~\cite{noauthor_provenance_nodate} to recover source commit and CI/CD workflows to recover build commands. 
However, they are tailored to Maven and do not establish whether the recovered metadata is sufficient for artifact-level verification across source-oriented ecosystems.

Google's OSS Rebuild~\cite{google_oss_rebuild} extends rebuild efforts to Crates.io, npm, and PyPI, using similar metadata-recovery heuristics. However, it focuses on supported popular packages and reports using human augmentation when automation fails~\cite{noauthor_introducing_2025}. 
As a result, it remains unclear how far artifacts in decentralized-build ecosystems can be reproduced using only metadata available to an independent verifier.

\section{Knowledge Gaps and Research Questions}
\label{sec:knowledge-gaps}

Verifiable builds promise to mitigate software supply-chain attacks that exploit compromised build and release pipelines.
However, existing verifiable-build systems either operate in ecosystems with centralized build infrastructure, where rebuild metadata is recorded during build, or rely on human intervention to recover missing metadata in decentralized ecosystems.
As a result, we do not yet know whether artifacts in decentralized-build ecosystems can be verified automatically.

This paper asks: \textit{To what extent can published artifacts in decentralized-build package ecosystems be independently verified against their source without human intervention?}
We structure this question into four research questions:

\begin{description}
\JD{RQ1 includes the ablation on metadata extraction and whatever other knobs you've got}
\JD{Does combining strategies produce better results, or is one strategy dominant}

\item[\textbf{RQ1.}] \textit{Verifiability results:} What proportion of registry artifacts can be verified against their source?

\item[\textbf{RQ2.}] \textit{Root causes:} What are the root causes of artifact verifiability failures?

\item[\textbf{RQ3.}] \textit{Factor analysis:} How is verifiability affected by artifact characteristics and development choices?


\item[\textbf{RQ4.}] \textit{Comparison:} How does \toolname compare with state-of-practice rebuild systems?
\end{description}

\section{Models for Independent Artifact Verifiability}
\label{sec:verifiability-model}

We model two aspects of independent artifact verification:
  a \textit{verifier model}, to describe the capabilities of an independent verifier,
  and
  an \textit{artifact comparison model}, describing how to compare published vs. rebuilt artifacts.

\JD{Three design choices: (1) How you recover information, and (2) How you build, and (3) How you compare. Maybe a good opportunity for a checks-and-exes or otherwise comparison table placed here in the Overview. Then A, B, and C of this section cover each topic.}

\subsection{Verifier Model}
\label{subsec:verifier-model}

We model a \textbf{verifier} as an independent entity that validates a published artifact using only registry-exposed or registry-derived information, without access to the original build machine, private CI/CD state, or undisclosed secrets. Examples include registries performing publication-time checks and downstream users validating artifacts before installation.

The verifier makes two design choices: how to recover rebuild metadata and how to rebuild the artifact. 
Our model focuses on fully-automatable and generalizable metadata recovery.
It recovers metadata from three complementary sources: the published artifact, the registry-linked source repository, and registry-linked provenance when available. 
These sources provide evidence such as embedded source information, build-tool lock files, tags, commit history, manifests, source commits, and CI/CD workflows. 
We restrict recovery to these sources because they are available to independent verifiers and can scale across ecosystems and packages.

For rebuilding, our model adopts an independent builder approach~\cite{lamb_reproducible_2022}. 
Instead of replaying the original CI/CD workflow, the verifier uses ecosystem-level canonical build and packaging commands, such as \texttt{npm pack} for npm and \texttt{python -m build} for PyPI. 
This reduces dependence on potentially compromised workflow steps and tests whether artifacts can be reproduced from durable source and build metadata. 
However, it can fail when packages rely on custom workflow steps or package-specific build logic.

Unlike prior systems tailored to specific ecosystems or package sets, such as BuildGen for Maven~\cite{hassanshahiUnlockingReproducibilityAutomating2025} and OSS-Rebuild~\cite{google_oss_rebuild}, our model intentionally restricts the verifier to ecosystem-level, registry-derived evidence to measure how far independent artifact verification can scale across decentralized-build package ecosystems.

\subsection{Artifact Comparison Model}
\label{subsec:artifact-comparison-model}

Artifact verification requires a criterion for comparing the published artifact with the independently rebuilt artifact. 
Package artifacts in decentralized-build ecosystems are mostly structured archives with inspectable contents, rather than flat byte sequences as modeled by prior work~\cite{dietrich_levels_2025}.
As a result, we define an \textbf{artifact comparison model} that uses tiered equivalence levels over package contents and metadata.

\begin{itemize}
\item \textit{L1: byte-identical.} The published and rebuilt artifacts are byte-for-byte identical. This is the strongest level and matches the standard definition of reproducibility.

\item \textit{L2: content-identical.} The artifacts contain the same files with identical contents after normalizing archive-level metadata, such as timestamps, file ordering, compression parameters, ownership fields, and permissions.

\item \textit{L3: code-identical.} The artifacts match after excluding or normalizing known non-executable content differences, such as documentation, generated package metadata, runtime-irrelevant lock files, or non-executable fields in build definition files.

\item \textit{L4: behavioral equivalence.} The artifacts differ syntactically, but source-code, binary, or semantic analysis shows that the differences do not affect execution. We treat L4 as out of scope because such analyses may be unavailable, incomplete, or unsound for arbitrary package artifacts.
\end{itemize}

Unlike prior systems that report only L1 (bit-identical)~\cite{google_oss_rebuild} or rebuild success~\cite{hassanshahiUnlockingReproducibilityAutomating2025}, this model makes artifact comparison explicit and reports outcomes for multiple equivalence levels.

\section{Design and Implementation of Artifact Verification Pipeline (\toolname)}
\label{sec:toolname-design}

\begin{figure*}
    \centering
    \includegraphics[width=0.90\linewidth]{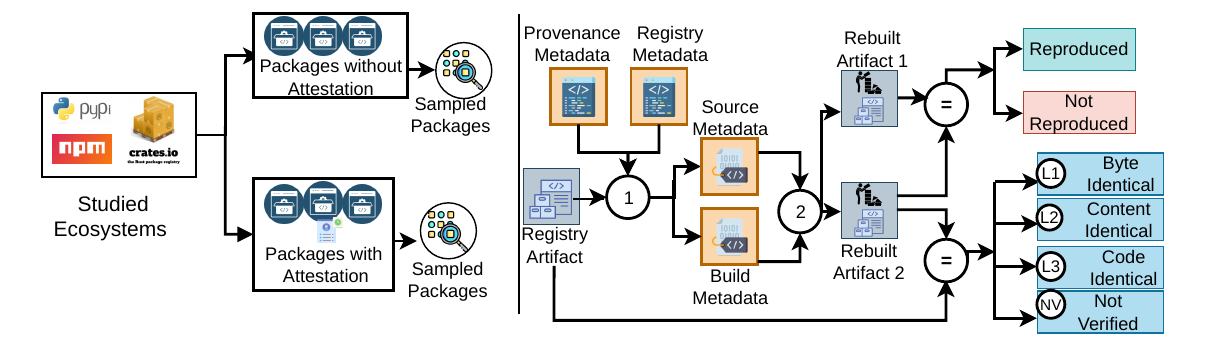}
    \caption{
    Methodology Overview and Artifact Verification Pipeline Design and Implementation.
   }
    \label{fig:method_overview}
\end{figure*}

We implement \toolname from the models in \cref{sec:verifiability-model} for four decentralized-build ecosystems: Crates.io, npm, PyPI, and RubyGems. As shown in \cref{fig:method_overview}, it operates in three stages: recovering rebuild metadata, rebuilding the artifact, and comparing the rebuild with the published artifact.

\subsection{Stage 1: Recover Rebuild Metadata}

The first stage recovers the metadata needed to rebuild an artifact:
  \emph{source metadata}, to identify the source code to rebuild;
  and
  \emph{build metadata}, to identify the tools, dependencies, timestamps, and commands needed to reconstruct the artifact.

\subsubsection{Recovering Source Metadata}
\label{subsubsec:stage-1-source-recovery}

Source metadata consists of the repository URL, source commit, and package subpath.
\toolname recovers this metadata from three sources, in order of preference: provenance attestations, embedded source metadata, and heuristics.

For artifacts with provenance metadata, such as attestations (npm~\cite{npm-provenance}, PyPI~\cite{pep740}, RubyGems) or trusted publishing information (Crates.io~\cite{crates-trusted-publishing}), \toolname extracts the repository URL, commit hash, and GitHub Actions workflow URI from the provenance attestation or trusted publishing information present in the artifact's registry metadata.
In addition, for Crates.io packages with artifact-embedded source metadata (\eg \texttt{.cargo\_vcs\_info.json}~\cite{cargo-package}), \toolname extracts the commit and package subpath from the embedded file.

When these sources are unavailable, \toolname combines heuristics from prior work~\cite{hassanshahiUnlockingReproducibilityAutomating2025, google_oss_rebuild, keshani_aroma_2024}.
It recovers repository URLs from registry fields such as project URL, homepage, and description~\cite{gao_pyradar_2024, vu_lastpymile_2021}, then normalizes and validates candidate URLs.
It recovers commits using three ordered strategies: \textit{tag matching} against the artifact version; \textit{version-string search} in commits within \texttt{[-7d, +2d]} of artifact upload time; and, as a final fallback, the \textit{latest default-branch commit} before upload.
Finally, it identifies the package subpath by locating the directory whose package definition matches the artifact name, using lexical matching when no exact match exists.

\subsubsection{Recovering Build Metadata}

Build metadata consists of the tools, versions, dependencies, commands, and timestamps needed to rebuild the artifact. \toolname recovers it from ecosystem-specific files in the artifact and source tree---for example, the build backend, its version, and dependencies from PyPI's \texttt{pyproject.toml}, or the package manager from npm's \texttt{package.json} and lockfiles---and then constructs the build and packaging commands from the ecosystem and recovered build tool.


Finally, to reduce non-determinism from file modification times, \toolname uses the published artifact's uniform modification time as the build timestamp when one exists, and otherwise the recovered source-commit timestamp, following reproducible-builds guidance~\cite{reproducible-builds-sde}.

\subsection{Stage 2: Rebuild the Artifact}

The second stage uses the recovered metadata to produce an independent rebuild.
Following our independent verifier model (\cref{subsec:verifier-model}), \toolname uses ecosystem-default build and packaging commands to rebuild the artifact.

AVP first prepares up to three variants of the checked-out source tree: without Git metadata, with Git metadata, and with Git metadata plus build-tool \path{pretend-version} variables, to accommodate tools that derive versions or timestamps from version control. 
It then installs the recovered build tools and dependencies, sets \path{SOURCE_DATE_EPOCH} to the recovered timestamp, applies tool-specific setup (e.g., Corepack for pnpm), and runs the recovered build and packaging commands.
\subsection{Stage 3: Compare Artifacts}
\label{subsec:stage-3}

Stage~3 compares each rebuilt artifact with the published one and reports the strongest level reached. L1 compares artifact hashes; L2 compares per-file content hashes while allowing archive-metadata differences; L3 checks that residual differences fall within a small, fixed catalog of non-executable metadata files and fields built from ecosystem documentation (e.g., PyPI's \texttt{RECORD}, \texttt{METADATA}, \texttt{WHEEL}, \texttt{PKG-INFO}, and generated \texttt{\_version.py} files, npm's \texttt{package.json} metadata fields, and Cargo's \texttt{.cargo\_vcs\_info.json}, and \texttt{Cargo.lock}). Artifacts reaching none are not verified; we do not operationalize L4.

\section{Artifact Verifiability Methodology}
\label{sec:methodology}

This section describes our methodology for answering the research questions in \cref{sec:knowledge-gaps}.

\subsection{Experimental Setup}
\label{subsec:experimental-setup}

\subsubsection{Package Ecosystem Selection}
We study four source-oriented ecosystems: Crates.io, npm, PyPI, and RubyGems. These are among the largest programming-language ecosystems and frequent targets
of supply-chain attacks~\cite{ohm_backstabbers_2020}; they also match the source-oriented ecosystems studied by recent reproducible-build work~\cite{benedetti_empirical_2025} and supported by OSS Rebuild~\cite{google_oss_rebuild}. 

\subsubsection{Dataset Construction and Sampling}
\label{subsubsec:setup-artifact-selection}
For each ecosystem, we enumerate packages from the registry and build a package-level dataset of metadata, source-repository information, and provenance-attestation status. We recover repository URLs from registry fields such as project URL, homepage, and description, following prior
work~\cite{gao_pyradar_2024}, and query each ecosystem's provenance API for whether the latest artifact is attested.

\begin{figure}[t]
\centering
\includegraphics[width=0.8\linewidth]{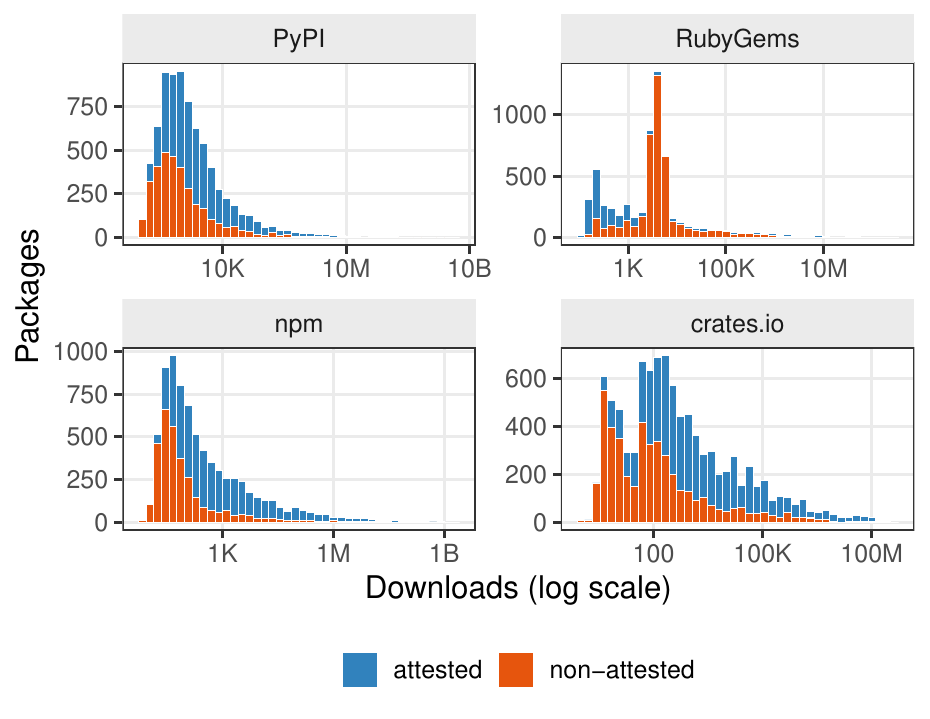}
\caption{Download distribution of the sampled packages per ecosystem.}
\label{fig:dataset-characteristics}
\end{figure}

Two properties of the dataset drive our sampling. Attestation adoption is low (2.0-6.6\%), so a random ecosystem-level sample would contain too few attested artifacts to evaluate attestations. Many packages lack valid repository URLs (14.7-42.4\%), which we report as an ecosystem-level source-recovery limitation. We exclude packages with no registry-linked public repository candidate from the rebuild experiment; however, \toolname may still fail source recovery for sampled packages when the linked repository is unavailable, invalid, lacks a matching package subpath, or no source commit can be recovered. We therefore sample within attested and non-attested strata separately, drawing up to 4,500 packages per stratum, above the sample needed for a 95\% confidence level and 1.5\% margin of error under the conservative infinite, population assumption of prior work~\cite{benedetti_empirical_2025}. For crates.io and RubyGems we take the full attested set---5{,}206 and 1{,}799 packages, respectively---since RubyGems has fewer than 4{,}500 attested packages in total, and for crates.io, we use all attested crates rather than subsampling. For each sampled package we select its most recent artifact as of May 2026, yielding 34{,}005 artifacts (\cref{tab:dataset-and-completion}); \Cref{fig:dataset-characteristics} shows the
download-rate distribution.

\subsubsection{Experiment environment}
\label{subsubsec:rebuild-env}

We run all rebuilds and measurements on an IBM Cloud bare-metal server profile: 48 physical cores, 384\, GB RAM, and a 960\, GB local NVMe disk. The server runs Ubuntu Linux 24.04 LTS (x86-64).

\subsection{RQ1: Proportion of Verified Software Packages}

RQ1 measures the proportion of packages in each ecosystem that can be independently verified under the models in \cref{sec:verifiability-model}. For each selected artifact, we run \toolname to recover rebuild metadata, reconstruct the build environment, rebuild the package twice, and compare the rebuilt artifacts against the published one. We run three experiments: (i) the non-attested ecosystem sample with heuristic metadata recovery, (ii) the attested
sample with heuristic source recovery, and (iii) the attested sample with attestation-based recovery. The latter two isolate the contribution of provenance attestations to verifiability.

We report two metrics per experiment: the \textit{completion rate} (the fraction of artifacts for which \toolname produces a comparison result, together with the reasons for pre-comparison failures), and the \textit{equivalence rate} (the fraction reaching each level). We consider an artifact \textit{verified} at L1, L2, or L3 (\cref{subsec:artifact-comparison-model}).

\subsection{RQ2: Root Causes of Non-L1 Verifiability Failures}

For artifacts that did not reach L1, we qualitatively analyzed the observed differences and their causes. We restricted this to artifacts with provenance metadata, whose attestations or trusted-publishing records identify the source commit and CI/CD workflow, enabling inspection and partial reconstruction of the original build.

In the first stage, two authors independently open-coded 48 randomly sampled non-L1 artifacts: for each, they examined the \texttt{diffoscope}~\cite{diffoscope} report to identify observable differences, then inspected the source commit and GitHub
Actions workflow to infer the underlying source-state, build-process, or environment cause (30-60 minutes per artifact). They reconciled their findings into a shared taxonomy and codebook defining each symptom, its possible causes, and ecosystem-specific manifestations.

In the second stage, we used a frontier coding agent to scale the analysis, requiring experimental validation for each proposed cause: the agent corrected the hypothesized cause in a Dockerfile matching our rebuild environment and
checked whether the artifact's equivalence level improved, with a second agent reviewing each classification. Applied over three phases to saturation, this covered 134 further artifacts (35 from Crates/npm/PyPI and 29 from RubyGems). We report the resulting causes
and symptoms, grouped by the rebuild metadata they affect, with frequencies across ecosystems.

\begin{table}[t]
\caption{Dataset and pipeline completion per ecosystem. Completion rate and failure reasons are \% of sampled artifacts.}
\label{tab:dataset-and-completion}
\centering
\begin{tabular}{lrrrr}
\toprule
 & Crates.io & npm & PyPI & RubyGems \\
\midrule
Sampled: attested       & 5{,}206 & 4{,}500 & 4{,}500 & 1{,}799 \\
Sampled: non-attested   & 4{,}500 & 4{,}500 & 4{,}500 & 4{,}500 \\
Total sampled           & 9{,}706 & 9{,}000 & 9{,}000 & 6{,}299 \\
\midrule
\;Completion rate         & 83.4\%  & 61.6\%  & 74.4\%  & 74.9\%  \\
\;\;\;\;No source/repo        & 4.1\%   & 15.5\%  & 13.4\%  & 15.8\%  \\
\;\;\;\;Clone/download        & 9.1\%   & 2.5\%   & 0.4\%   & 2.1\%   \\
\;\;\;\;Build/install         & 3.4\%   & 20.4\%  & 11.2\%  & 7.3\%   \\
\;\;\;\;Timeout/other         & 0.0\%   & 0.0\%   & 0.6\%   & 0.0\%   \\
\bottomrule
\end{tabular}
\end{table}

\subsection{RQ3: Impact of Artifact Characteristics and Development Choices}

RQ3 tests whether six factors are associated with verifiability: artifact size, download rate, and age (artifact characteristics), and build tool, native code,
and monorepo layout (development choices). The last three capture development choices that may complicate rebuilding: monorepos can obscure package subpaths, native extensions can depend on platform-specific environments, and build tools may introduce different sources of nondeterminism.
We stratify continuous factors into quartiles and group categorical factors, comparing verifiability across bins.



\subsection{RQ4: \toolname vs State-of-Practice Rebuild Platforms}
\label{subsec:method-rq5}

RQ4 compares \toolname with OSS-Rebuild, Google's open-source platform that infers build definitions, rebuilds artifacts, and produces rebuild attestations for popular packages. We select it as open source, targeting our ecosystems, and operated at scale. We compare on Crates.io, npm, and PyPI, excluding RubyGems, which OSS-Rebuild does not support.

We draw 600 artifacts: 200 per ecosystem, split into two strata of 100 by OSS-Rebuild coverage. The supported stratum is sampled from packages OSS-Rebuild already tracks; the unsupported stratum from each ecosystem's most popular
packages by download count, excluding any already supported. Both strata are thus widely-used software differing only in coverage, which separates how the platforms compare where OSS-Rebuild is tuned from how they compare on the popular
long tail it does not support. This frame is independent of the attestation-stratified dataset used in our other RQs.

We rebuild each artifact with both platforms from the same published version: OSS-Rebuild via its tool, which infers a build definition, builds hermetically, and reports whether the result matches; \toolname via our pipeline
(\cref{fig:method_overview}) on the same artifact and recovered source commit. Because OSS-Rebuild reports a binary match while \toolname reports graded levels (\cref{subsec:artifact-comparison-model}), we count an \toolname artifact as
verified at L1--L3, normalizing the same non-semantic differences OSS-Rebuild normalizes when deciding a match. We report, per ecosystem and stratum, each platform's verification rate and their agreement (verified by both, \toolname only, OSS-Rebuild only, or neither); for artifacts verified by exactly one, we analyze the design choices: source recovery, build-environment inference,
native-code handling, that explain the divergence.

\begin{figure}
\centering
\includegraphics[width=1.05\linewidth]{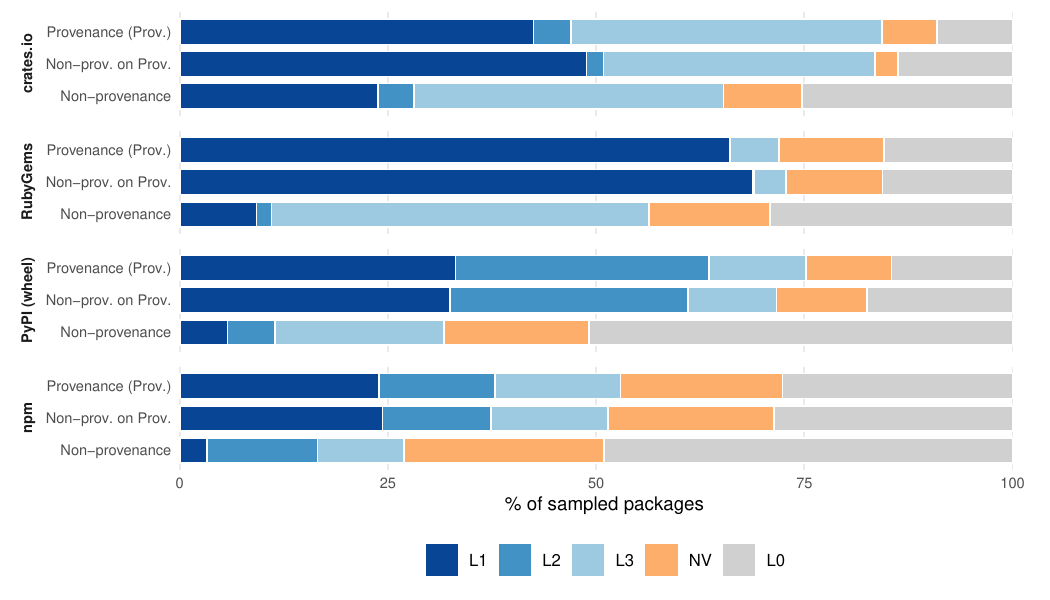}
\caption{Artifact equivalence by ecosystem and source-recovery strategy, over artifacts the pipeline could compare (completion rates in \Cref{tab:dataset-and-completion}).}
\label{fig:rq1-equivalence}
\end{figure}

\section{Results}
\label{sec:results}

This section presents the results of our empirical study.

\subsection{RQ1: Ecosystem Verifiability Results}
\label{subsec:results-rq1}

\Cref{fig:rq1-equivalence} reports artifact equivalence across ecosystems and source-recovery strategies, while \Cref{tab:dataset-and-completion} reports pipeline completion and failure causes. We separate these outcomes: \emph{completion} measures whether \toolname recovers the source and rebuilds the artifact, while \emph{equivalence} measures how closely completed rebuilds match the published artifact.

\myparagraph{Internal reproducibility}
We first confirm that \toolname's rebuilds are deterministic by building each package twice.
These rebuilds were byte-identical for 93.0--100.0\% of artifacts across ecosystems and content-identical for 94.8--100.0\%, exceeding reported reproducibility rates for these ecosystems~\cite{benedetti_empirical_2025}. 

\myparagraph{Heuristic recovery}
For artifacts without provenance, \toolname recovers the source commit heuristically~(\Cref{subsubsec:stage-1-source-recovery}). Pipeline completion is 74.7\%, 50.9\%, 49.1\%, and 70.9\% for Crates.io, npm, PyPI, and RubyGems, respectively~(\Cref{tab:dataset-and-completion}). The dominant failure is source-commit recovery, a limitation shared by prior rebuild systems and empirical studies that rely on similar heuristics~\cite{imtiaz_dependencies_reviewed_2023,google_oss_rebuild,hassanshahiUnlockingReproducibilityAutomating2025,benedetti_empirical_2025,gao_pyradar_2024,vu_lastpymile_2021,goswami_npm_2020}. Among artifacts that complete, 87.4\% of Crates.io artifacts verify at L1--L3, compared with 52.9\% for npm, 64.6\% for PyPI, and 79.4\% for RubyGems. Crates.io performs best on both completion and equivalence because crate artifacts embed source metadata that directly supports source recovery.

\myparagraph{Provenance-based recovery}
For provenance-bearing packages, \toolname recovers the source commit directly from npm and PyPI attestations, Crates.io Trusted Publishing metadata, or RubyGems attestations. This improves both completion and equivalence. For example, npm completion rises from 50.9\% to 72.4\%, and equivalence among completed artifacts rises from 52.9\% to 73.2\%. RubyGems similarly improves from 70.9\% to 84.5\% completion and from 79.4\% to 85.1\% equivalence, with comparable gains for Crates.io and PyPI~(\Cref{fig:rq1-equivalence}). Once the source commit is known, remaining failures are dominated by build-execution errors and missing build dependencies rather than source recovery.

\myparagraph{Isolating the effect of provenance metadata}
Provenance-bearing packages may also be more popular, active, or mature, so their higher verifiability may not be caused by provenance metadata alone. To isolate the metadata effect, we hold the package set fixed and compare provenance-based recovery with heuristic recovery on the same provenance-bearing packages. Measured as the overall verified fraction of the sample, the direct gain from recorded provenance is small: 0.8, 1.5, 3.6, and $-0.9$ percentage points for Crates.io, npm, PyPI, and RubyGems, respectively. In contrast, the gap between heuristic recovery on provenance-bearing packages and heuristic recovery on the general sample is much larger: 18.2, 24.5, 39.9, and 16.5 points. Thus, most of the verifiability advantage of provenance-bearing packages reflects the build and release practices of adopting projects, not the provenance metadata itself.

\begin{rqanswer}{RQ1 Takeaway}
Verifiability is low and uneven across ecosystems: among completed rebuilds, 52.9--87.4\% verify at L1--L3, and source-commit recovery is the dominant pipeline failure. Provenance improves source recovery and raises apparent verifiability, but holding the package set fixed shows that the recorded metadata contributes at most 3.6 percentage points. Provenance makes source recovery reliable; it does not, by itself, make artifacts verifiable.
\end{rqanswer}

\subsection{RQ2: Root Causes of Non-L1 Verifiability Failures}
\label{subsec:results-rq2}

\newcolumntype{Y}{>{\raggedright\arraybackslash}X}

\begin{table*}[t]
\centering
\scriptsize
\setlength{\tabcolsep}{3.2pt}
\renewcommand{\arraystretch}{1.12}
\caption{Root causes of artifact verifiability failures and the symptoms they produce. We analyzed 134 artifacts, resulting in 185 root cause identifications.}
\label{tab:failure-root-causes}
\begin{tabularx}{\textwidth}{@{}p{0.06\textwidth} p{0.24\textwidth} Y p{0.18\textwidth} c@{}}
\toprule
\textbf{ID} & \textbf{Root Cause} & \textbf{Failure Symptoms} & \textbf{Reconstruction Gap} & \textbf{Count} \\
\midrule
RC01 & Build tool or dependency version differences & Dependency or file version differences in metadata files; version fields of build tools and/or dependencies in manifest or lock files differ; generated files differ because build tool or dependency versions differ. & Environment reconstruction & 47 \\
\midrule
RC02 & CI/CD source code modification & Original artifact contains files or metadata modified by CI/CD workflow steps; package version fields, version-dependent paths, or pre-processed source files differ from the source commit. & Source-state reconstruction & 38 \\
\midrule
RC03 & Generated files or artifacts produced by CI/CD workflow & Generated files or intermediate build outputs such as compiled files or downloaded assets exist in the original artifact but are missing in the rebuilt artifact; files in the original artifact do not exist in the source repository; published artifact represents generated package content. & Build-process reconstruction & 28 \\
\midrule
RC04 & Original timestamp not fixed & Timestamp fields in archives differ. & Build-process reconstruction & 18 \\
\midrule
RC05 & Packaging directory differences & Files in original and rebuilt artifacts were packaged from different temporary build directories. & Build-process reconstruction & 14 \\
\midrule
RC06 & Source revision mismatch & There is a drift between the source commit that produced the artifact and the commit used for rebuilding. & Source-state reconstruction & 11 \\
\midrule
RC07 & Environment-dependent metadata in generated files or file names & Generated files contain local file paths from the build runner environment, or generated file names depend on host paths or directory names. & Environment reconstruction & 6 \\
\midrule
RC08 & Native artifact differences & Native binaries, wheel tags, repair metadata, records, or bundled native files differ. & Environment reconstruction & 5 \\
\midrule
RC09 & Package subpath differences & Files in the rebuilt artifact correspond to a different subdirectory or package in the source repository. & Source-state reconstruction & 4 \\
\midrule
RC10 & Git submodule materialization differences & Files from a Git submodule exist in one artifact but are missing in the other. & Source-state reconstruction & 4 \\
\midrule
RC11 & External VCS source inputs in artifacts & Additional files in original artifacts are VCS-tracked external source payloads, such as Git LFS project files, that were not materialized in the rebuild. & Source-state reconstruction & 4 \\
\midrule
RC12 & Build output contains unstable or unordered inputs & Generated tables, manifests, caches, source files, file names, directories, source maps, or embedded IDs differ across rebuilds. & Build-process reconstruction & 4 \\
\midrule
RC13 & Archive container encoding not fixed & File contents, paths, modes, and mtimes match, but archive SHA differs. & Build-process reconstruction & 1 \\
\midrule
RC14 & VCS line-ending normalization differences & Shared text files differ only by CRLF/LF normalization. & Source-state reconstruction & 1 \\
\bottomrule
\end{tabularx}
\end{table*}

\Cref{tab:failure-root-causes} summarizes the observed root causes, their symptoms, and the reconstruction gaps they expose. For space, we discuss three representative causes.

\myparagraph{Build tool or dependency version differences}
The most common source of divergence was version drift in build tools or build dependencies. 
These differences produced either L3 metadata differences, such as changed lock files, or content differences in files generated by the build toolchain. 
For example, \texttt{tidesurf}~0.2.1 was originally built with Cython~v3.2.3, while our rebuild resolved to Cython~v3.2.6, producing different generated C files.
In such cases, the relevant versions were not pinned in the source manifests, allowing dependency resolution to drift between the original build and the rebuild.

\myparagraph{Generated files produced by CI/CD workflow}
The second most common source of divergence was intermediate files generated by CI/CD workflow steps but not by the canonical build process used in our verifier model. For example, the npm \texttt{lingui-swc}~0.6.0 workflow ran \texttt{yarn build:ts} and \texttt{napi prepublish}, generating files under \texttt{dist/} that were included in the published artifact but absent from our rebuild.
These cases show that artifact verification can fail when build-relevant workflow steps are not captured by registry-exposed metadata or ecosystem-standard build interfaces; prior work similarly finds published packages containing files absent from their source repositories~\cite{vu_lastpymile_2021}.

One might expect that \emph{replaying} the original workflow would close this gap. We evaluated this directly in a pilot: for 105 Crates.io and 105 npm artifacts sampled per outcome bucket, we re-ran each package's real release workflow using \texttt{act}~\cite{nektos_act} and compared the result to the published artifact.\footnote{We excluded other ecosystems after poor results on Crates.io and npm.} Replay improved at most 4\% of artifacts (Crates.io 1/105, npm 4/105) and closed essentially no existing divergences: every npm gain was coverage on a previously failed build rather than a match, and the sole genuine recovery (\texttt{ruckup}~0.9.1, which turned a mismatch into an exact L1 match) succeeded only because the workflow pinned a build-toolchain version our heuristic had guessed wrong. Replay is also fragile: even trivially reproducing artifacts fail to build under \texttt{act} in roughly 9 of 10 cases. Workflow replay therefore does not scale as a verification strategy; the one real gain points instead at build-environment pinning, which we return to in \Cref{subsec:discussion-verifiability-metadata}.

\myparagraph{Source revision mismatch}
Source revision mismatches occur when provenance metadata records a different commit from the one that produced the published artifact.  For example, the Crates.io \texttt{aviutl2}~0.27.0 crate pointed to commit \texttt{7dcd140} and tag \texttt{0.27.0}, while trusted publishing metadata recorded commit \texttt{b2d6bc16}; rebuilding from the recorded commit produced an L1-equivalent artifact. These mismatches often arose from semantic-release-style workflows that modify source metadata, commit the changes, and then publish the artifact; provenance records the triggering commit, not necessarily the final source commit used to produce the artifact.

\begin{rqanswer}{RQ2 Takeaway}
61\% of analyzed divergences were caused by build tool or dependency version drift, workflow-generated intermediate files, or source modifications introduced by CI/CD steps. Improving artifact verifiability requires infrastructure that records build tool versions, source-state changes, and build-relevant workflow steps so that verifiers can accurately reconstruct the source state, environment, and build process used to produce the artifact.
\end{rqanswer}

\begin{figure}[t]
\centering
\includegraphics[width=0.9\linewidth]{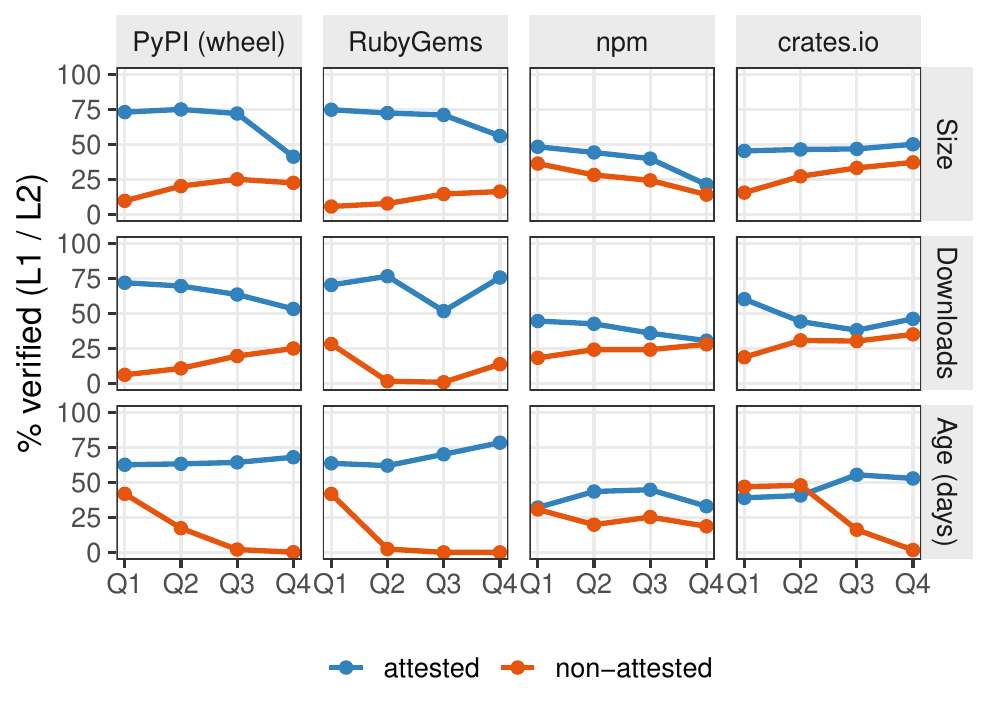}
\caption{Verification rate (\% reproduced at L1/L2) by within-ecosystem quartile (Q1 lowest, Q4 highest) of artifact size, downloads, and age, for attested vs.\ non-attested packages across four ecosystems.}
\label{fig:rq3-continuous-factors}
\end{figure}

\subsection{RQ3: Artifact Characteristics and Development Choices}
\label{subsec:results-rq3}

We consider two classes of factors that may explain verifiability outcomes: artifact characteristics, such as size, age, and popularity, and development choices, such as build backend, native code, and repository layout. \Cref{fig:rq3-continuous-factors} shows that artifact characteristics have limited explanatory power: verifiability declines only slightly with artifact size, shows little relationship with popularity, and the apparent effect of release age largely disappears once attestation is controlled for. In contrast, development choices have a much stronger effect. Across ecosystems, verifiability is primarily shaped by how deterministically the build and packaging process reconstructs the published artifact.

\myparagraph{Packaging determinism matters more than closeness to source}
Crates.io performs consistently well because crates are published as source tarballs through a standardized packaging process, with remaining mismatches mostly caused by embedded build timestamps~(\Cref{subsec:results-rq2}). However, being close to source is not sufficient. Package artifacts can diverge from their repositories even without compilation, because packaging can rewrite metadata, normalize files, or apply preprocessing~\cite{vu_lastpymile_2021}. npm illustrates this effect: \texttt{pack-only} packages verify less often than bundled or transpiled packages, even though the latter transform the source more heavily. The difference is packaging determinism: stable generated outputs can be easier to reproduce than source-like artifacts produced by nondeterministic packaging steps.

\myparagraph{PyPI has high-verifiability backends \& low-verifiability paths}
PyPI shows the widest spread across build backends. At the high end, \texttt{hatchling} reproduces many wheels at L1--L2, requiring little normalization. Other backends, including \texttt{pdm-backend}, \texttt{poetry-core}, and \texttt{setuptools}, reach comparable total verification rates mainly through L3 normalization, indicating that their outputs often differ in metadata, timestamps, or ordering rather than executable content. At the low end, wheels built through legacy \texttt{setup.py bdist\_wheel} paths or packages without a declared backend rarely verify. This spread helps explain why PyPI contains some highly verifiable build configurations while remaining difficult to verify overall.

\myparagraph{Native compilation reduces verifiability}
Native compilation substantially lowers verification rates in ecosystems that ship built artifacts. On PyPI, wheels with compiled extensions, such as those built with \texttt{maturin} or \texttt{scikit-build-core}, verify far less often than pure-Python wheels; on npm, \texttt{node-gyp} modules similarly sit near the bottom of the ecosystem. These packages depend on platform-specific compilers, flags, and toolchain behavior, introducing variation that package-level normalization cannot remove~\cite{malkaReproducibilityBuildEnvironments2024,randrianainaOptionsMatterDocumenting2024}. Native packaging therefore remains a major barrier to independent artifact verification.

\myparagraph{Monorepos mainly affect source recovery}
Repository layout has little effect once the correct package directory is recovered. PyPI packages built from monorepo subdirectories verify at the same rate as root-level packages, suggesting that monorepos primarily create a source-recovery challenge rather than a rebuild or comparison challenge. Their impact therefore appears in pipeline completion failures~(\Cref{tab:dataset-and-completion}), not in equivalence outcomes among completed rebuilds.

\begin{figure}[t]
\centering
\includegraphics[width=\linewidth]{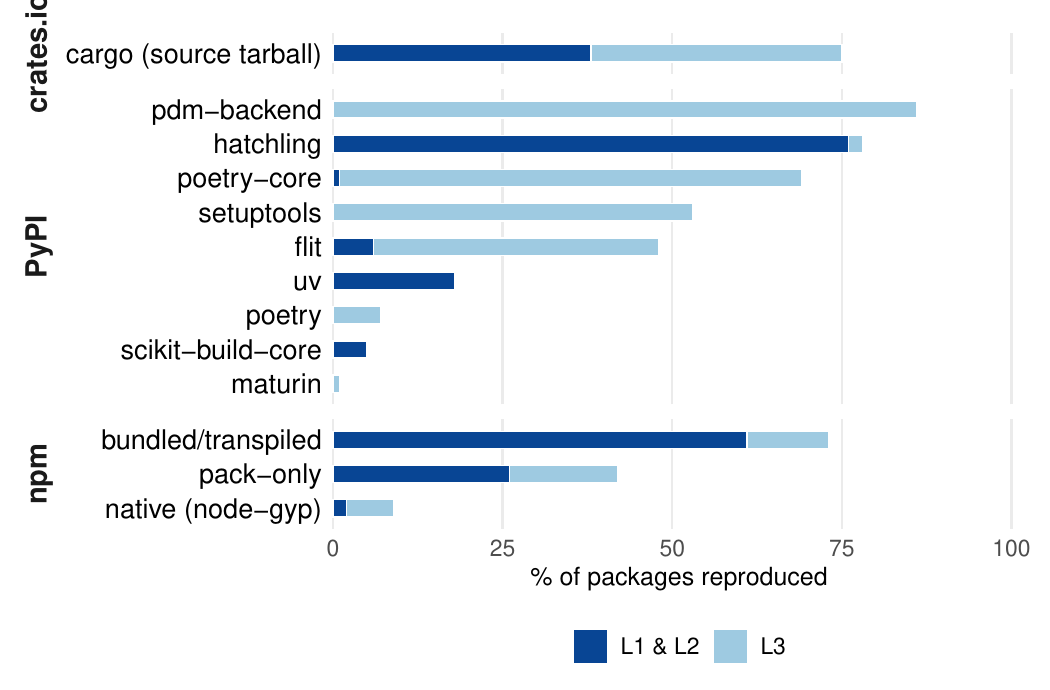}
\caption{Packages verified (\%) by build backend or packaging strategy, grouped by ecosystem and split into L1--L2 and L3 matches.}
\label{fig:rq3-dev-categorical}
\end{figure}

\begin{rqanswer}{RQ3 Takeaway}
Verifiability is shaped more by how an artifact is built than by what the artifact is. Size, age, and popularity have limited effects, while deterministic packaging, declarative build backends, and avoidance of native compilation strongly improve verification outcomes. Monorepos mainly affect source recovery rather than rebuild equivalence, so the most important levers for improving verifiability are build- and packaging-side changes.
\end{rqanswer}

\subsection{RQ4: \toolname vs.\ OSS-Rebuild}
\label{subsec:results-rq4}

Across the 600 artifacts, \toolname verifies 331 (55\%) versus 158 (26\%) for
OSS-Rebuild---more than twice as many. \Cref{tab:rq4-comparison} provides the details. 
\toolname attains the higher verification rate in four of the six strata, ties in one (PyPI unsupported), and is exceeded only on supported npm packages.

\myparagraph{\toolname generalizes to the long tail; OSS-Rebuild does not}
The starkest gap is the unsupported long tail: on unsupported crates.io, OSS-Rebuild verifies \emph{none} (0\%) against \toolname's 74\%, and 74\%, 22\%, and 6\% of unsupported crates, npm, and PyPI artifacts respectively are verified by \toolname alone. OSS-Rebuild depends on curated, per-package build definitions, so outside its supported set it often has no recipe to run---most visibly on crates.io, where it rebuilt no unsupported package. \toolname instead recovers the published source commit and rebuilds from it, transferring to arbitrary packages without curation.
\begin{table}[t]
  \centering
  \caption{Verification by \toolname{} and OSS-Rebuild on 600 artifacts
  (100 per ecosystem/stratum).}
  \label{tab:rq4-comparison}
  \scriptsize
  \resizebox{0.9\columnwidth}{!}{%
  \begin{tabular}{@{}llrrrr@{}}
    \toprule
    Eco. & Strat. & \toolname-only & OSS-only & Both & None \\
    \midrule
    crates & supp.    & 61\% & 2\%  & 25\% & 12\% \\
    crates & un-supp. & 74\% & 0\%  & 0\%  & 26\% \\
    npm    & supp.    & 6\%  & 13\% & 33\% & 48\% \\
    npm    & un-supp. & 22\% & 13\% & 21\% & 44\% \\
    PyPI   & supp.    & 44\% & 4\%  & 22\% & 30\% \\
    PyPI   & un-supp. & 4\%  & 6\%  & 19\% & 71\% \\
    \midrule
    \multicolumn{2}{@{}l}{Overall} & 35\% & 6\% & 20\% & 39\% \\
    \bottomrule
  \end{tabular}%
  }
  \vspace{-0.5em}
\end{table}

\myparagraph{\toolname is competitive even where OSS-Rebuild is tuned}
Even on OSS-Rebuild's supported strata, \toolname verifies more crates.io (86\% vs.\ 27\%) and PyPI (66\% vs.\ 26\%) artifacts; OSS-Rebuild leads only on supported npm (46\% vs.\ 39\%), where its hand-maintained definitions reproduce bundled and transpiled JavaScript that \toolname's generic \texttt{npm pack} does not. The platforms are complementary, not redundant (\Cref{tab:rq4-comparison}): on supported crates.io they agree on only 25\% of artifacts while \toolname uniquely verifies a further 61\%, and OSS-Rebuild uniquely verifies only a small share anywhere (0--13\%). A non-trivial 12--71\% is verified by neither---native compilation, undeclared build backends, and unrecoverable source.

\begin{rqanswer}{RQ4 Takeaway}
\toolname verifies about twice as many artifacts as OSS-Rebuild overall and is the only platform that scales beyond a curated package set. It trades some per-package fidelity on heavily transformed artifacts (e.g., supported npm) for broad coverage that needs no per-package build definitions, making generality the decisive advantage for assessing artifact authenticity at scale.
\end{rqanswer}

\section{Threats to Validity}

\myparagraph{Construct and internal validity}
Our verifiability rates are lower bounds. We count an artifact as unverified when \toolname cannot recover its rebuild metadata or reproduce it with canonical build commands. Thus, the rates reflect both ecosystem verifiability and \toolname's verifier model, and environment reconstruction; a more complete verifier may verify more artifacts.

Our L3 results depend on a manually curated catalog of non-executable differences, built from ecosystem documentation and analysis of non-L1 cases. This catalog may be too restrictive or permissive, so we interpret L3 verification as evidence of source-to-artifact correspondence, not as proof that all normalized differences are security-neutral.

Our RQ2 root-cause analysis may also misattribute failures. We used an automated agent to scale manual coding and accepted hypothesized causes when corrective changes improved equivalence. Such changes may identify contributing causes rather than the only or true cause. Moreover, the taxonomy is derived from provenance-bearing artifacts, whose failures may differ from non-attested artifacts.

\myparagraph{External validity and reliability}
Our study evaluates only the most recent artifact of each package~(\cref{subsubsec:setup-artifact-selection}), so it does not capture version-to-version variation or temporal drift in tools, dependencies, and attestation adoption.

We restrict rebuilds to packages with recoverable public source repositories, excluding 14.7-42.4\%~\cite{PypiNpmGHAccessibility, tsakpinis2026investigatingnotablemetadatapractices}. Reported rates are therefore conditional on source recovery and may overestimate ecosystem-wide verifiability. Because we oversample attested packages, ecosystem rates are stratum-weighted rather than population-weighted, and we may miss packages with attestations outside registry-provided pathways.

Finally, our OSS-Rebuild comparison reflects one tool version and our operation of it, which may miss capabilities requiring additional configuration or expertise.

\section{Discussion}

Open-source software supply-chain standards recommend reproducible builds as a means of improving trust in distributed software artifacts.
We measured the gap between these recommendations and artifact verifiability in practice.

\subsection{Verifiability Requires Artifact-Bound Rebuild Metadata}
\label{subsec:discussion-verifiability-metadata}

\Cref{subsec:results-rq1} shows that end-to-end verification (over all sampled artifacts) ranges between 25\% and 80\% across ecosystems.
\Cref{subsec:results-rq2} further shows that many failures stem from unpinned build tools, CI/CD source modifications, and custom packaging.
Independent verifiability therefore requires more than deterministic builds: it needs durable, structured, artifact-bound or recoverable metadata that lets independent parties rebuild and compare the published artifact.

Centralized build ecosystems such as Debian, Arch Linux, and Nix partly address this problem by recording rebuild metadata through ecosystem-controlled build infrastructure and providing verifiers that consume them.
This model is difficult to apply directly to decentralized-build ecosystems where maintainers use diverse build tools, packaging conventions, dependency managers, and CI/CD workflows.
A more practical direction is for other centralized infrastructure, such as CI/CD workflow runners, to record rebuild metadata while executing workflow steps and embed in artifacts before publishing.
Our replay pilot (\Cref{subsec:results-rq2}) reinforces this: the one artifact recovered by re-running its workflow matched only because the workflow pinned the build-toolchain version our heuristic missed---it is the build environment, not the workflow logic, that must be recorded.
Additionally, developer practices that improve verifiability, such as pinning dependencies and including package preparation and build steps in package manifest or other metadata files, should be encouraged.

This also bears on supply-chain standards. Existing standards recommend reproducible builds~\cite{noauthor_securing_nodate}, but our results show reproducibility becomes actionable only when ecosystems also make recovering rebuild metadata feasible; standards should therefore add concrete guidance on which metadata to preserve, where verifiers obtain it, and how results are represented, informed by our RQ2 causes (\cref{subsec:results-rq2}).

\subsection{Deploying Verifiability in Practice}
\label{subsec:discussion-deploying-verifiability}

This paper reframes reproducible builds as a necessary but incomplete condition for software supply-chain integrity in decentralized-build ecosystems.
The practical barrier is whether an independent verifier can recover enough rebuild metadata to reproduce the published artifact. 
By identifying the metadata gaps and packaging practices that prevent verification, this study gives registries, standards bodies, and rebuild-tool developers concrete targets for making artifact verification practical at package-registry scale.

One challenge is how verifiability can be integrated into package lifecycles without immediate, ecosystem-wide enforcement. 
Registries are a natural deployment point---they already receive, validate, and distribute artifacts---and could adopt it incrementally: first publishing verification metadata and warnings, then prioritizing stronger equivalence levels for popular, critical, or security-sensitive packages. Downstream package managers and organizational policies can require a target equivalence level (\cref{subsec:artifact-comparison-model}) before installation to protect users.

Provenance attestations offer another path: they improve metadata recovery but attest only to the build process, not the artifact's authenticity~\cite{openssf_introducing_nodate}. They could instead distinguish \emph{workflow} provenance from \emph{package-build} provenance providing sufficient evidence to reproduce the artifact~\cite{noauthor_build_nodate}.

\section{Conclusion}
We studied whether artifacts in decentralized-build package ecosystems can be independently verified against their source, formalizing a registry-only verifier model and tiered artifact-comparison levels and implementing both in \toolname. Across 34{,}005 packages from Crates.io, npm, PyPI, and RubyGems, verifiability remains low and uneven: the barrier is not only build non-determinism but the difficulty of recovering the exact source revision, environment, and build process behind each artifact, and provenance attestations aid source recovery without yielding complete rebuild specifications. Compared with OSS-Rebuild, \toolname verifies about twice as many artifacts and covers a broader set of popular packages. Overall, artifact verifiability at registry scale requires durable, artifact-bound rebuild metadata.



\bibliographystyle{IEEEtran}
\clearpage
\bibliography{references}

@inproceedings{fourne_its_2023,
	location = {San Francisco, {CA}, {USA}},
	title = {It’s like flossing your teeth: On the Importance and Challenges of Reproducible Builds for Software Supply Chain Security},
	rights = {https://doi.org/10.15223/policy-009},
	isbn = {978-1-6654-9336-9},
	url = {https://ieeexplore.ieee.org/document/10179320/},
	doi = {10.1109/SP46215.2023.10179320},
	shorttitle = {It’s like flossing your teeth},
	eventtitle = {2023 {IEEE} Symposium on Security and Privacy ({SP})},
	pages = {1527--1544},
	booktitle = {2023 {IEEE} Symposium on Security and Privacy ({SP})},
	publisher = {{IEEE}},
	author = {Fourné, Marcel and Wermke, Dominik and Enck, William and Fahl, Sascha and Acar, Yasemin},
	urldate = {2026-03-24},
	date = {2023-05},
	year = {2023},
}

@article{lamb_reproducible_2022,
	title = {Reproducible Builds: Increasing the Integrity of Software Supply Chains},
	volume = {39},
	rights = {https://ieeexplore.ieee.org/Xplorehelp/downloads/license-information/{IEEE}.html},
	issn = {0740-7459, 1937-4194},
	url = {https://ieeexplore.ieee.org/document/9403390/},
	doi = {10.1109/MS.2021.3073045},
	shorttitle = {Reproducible Builds},
	pages = {62--70},
	number = {2},
	journaltitle = {{IEEE} Software},
	journal = {{IEEE} Software},
	shortjournal = {{IEEE} Softw.},
	author = {Lamb, Chris and Zacchiroli, Stefano},
	urldate = {2026-03-24},
	date = {2022-03},
	year = {2022},
}

@inproceedings{benedetti_empirical_2025,
	location = {Ottawa, {ON}, Canada},
	title = {An Empirical Study on Reproducible Packaging in Open-Source Ecosystems},
	rights = {https://doi.org/10.15223/policy-029},
	isbn = {979-8-3315-0569-1},
	url = {https://ieeexplore.ieee.org/document/11029905/},
	doi = {10.1109/ICSE55347.2025.00136},
	eventtitle = {2025 {IEEE}/{ACM} 47th International Conference on Software Engineering ({ICSE})},
	pages = {1052--1063},
	booktitle = {2025 {IEEE}/{ACM} 47th International Conference on Software Engineering ({ICSE})},
	publisher = {{IEEE}},
	author = {Benedetti, Giacomo and Solarin, Oreofe and Miller, Courtney and Tystahl, Greg and Enck, William and Kästner, Christian and Kapravelos, Alexandros and Merlo, Alessio and Verderame, Luca},
	urldate = {2026-03-23},
	date = {2025-04-26},
	year = {2025},
}

@inproceedings{PypiNpmGHAccessibility,
author = {Tsakpinis, Alexandros and Pretschner, Alexander},
title = {Analyzing the Accessibility of GitHub Repositories for PyPI and NPM Libraries},
year = {2024},
isbn = {9798400717017},
publisher = {Association for Computing Machinery},
address = {New York, NY, USA},
url = {https://doi.org/10.1145/3661167.3661231},
doi = {10.1145/3661167.3661231},
booktitle = {Proceedings of the 28th International Conference on Evaluation and Assessment in Software Engineering},
pages = {345–350},
numpages = {6},
location = {Salerno, Italy},
series = {EASE '24}
}

@misc{tsakpinis2026investigatingnotablemetadatapractices,
      title={Investigating Notable Metadata Practices in PyPI Libraries: An Empirical Study about Repository and Donation Platform URLs}, 
      author={Alexandros Tsakpinis and Nicolas Raube and Alexander Pretschner},
      year={2026},
      eprint={2601.15139},
      archivePrefix={arXiv},
      primaryClass={cs.SE},
      url={https://arxiv.org/abs/2601.15139}, 
}

@article{imtiaz_dependencies_reviewed_2023,
  author  = {Nasif Imtiaz and Laurie A. Williams},
  title   = {Are Your Dependencies Code Reviewed?: Measuring Code Review Coverage in Dependency Updates},
  journal = {IEEE Transactions on Software Engineering},
  volume  = {49},
  number  = {11},
  pages   = {4932--4945},
  year    = {2023},
  doi     = {10.1109/TSE.2023.3319509},
  url     = {https://doi.org/10.1109/TSE.2023.3319509}
}

@inproceedings{vu_lastpymile_2021,
author    = {Vu, Duc Ly and Pashchenko, Ivan and Massacci, Fabio and Plate, Henrik and Sabetta, Antonino},
title     = {LastPyMile: Identifying the Discrepancy between Sources and Packages},
booktitle = {Proceedings of the 29th ACM Joint European Software Engineering Conference and Symposium on the Foundations of Software Engineering},
series    = {ESEC/FSE '21},
year      = {2021},
publisher = {Association for Computing Machinery},
doi       = {10.1145/3468264.3468592},
url       = {https://doi.org/10.1145/3468264.3468592}
}

@inproceedings{goswami_npm_2020,
author    = {Goswami, Pronnoy and Gupta, Saksham and Li, Zhiyuan and Meng, Na and Yao, Daphne},
title     = {Investigating The Reproducibility of NPM Packages},
booktitle = {2020 IEEE International Conference on Software Maintenance and Evolution (ICSME)},
pages     = {677--681},
year      = {2020},
publisher = {IEEE},
doi       = {10.1109/ICSME46990.2020.00071},
url       = {https://doi.org/10.1109/ICSME46990.2020.00071}
}

@misc{npm-provenance,
author       = {{npm Docs}},
title        = {Generating Provenance Statements},
year         = {2026},
howpublished = {\url{https://docs.npmjs.com/generating-provenance-statements/}},
note         = {Accessed: 2026-06-08}
}

@misc{cargo-package,
author       = {{The Cargo Book}},
title        = {cargo-package(1)},
year         = {2026},
howpublished = {\url{https://doc.rust-lang.org/cargo/commands/cargo-package.html}},
note         = {Accessed: 2026-06-08}
}

@misc{crates-trusted-publishing,
author       = {{crates.io Docs}},
title        = {Trusted Publishing},
year         = {2026},
howpublished = {\url{https://crates.io/docs/trusted-publishing}},
note         = {Accessed: 2026-06-08}
}

@misc{google_oss_rebuild,
author       = {{Google}},
title        = {{OSS Rebuild}},
year         = {2025},
howpublished = {\url{https://github.com/google/oss-rebuild}},
note         = {Accessed: 2026-06-08}
}

@article{keshani_aroma_2024,
author  = {Keshani, Mehdi and Velican, Tudor-Gabriel and Bot, Gideon and Proksch, Sebastian},
title   = {AROMA: Automatic Reproduction of Maven Artifacts},
journal = {Proceedings of the ACM on Software Engineering},
volume  = {1},
number  = {FSE},
pages   = {836--858},
year    = {2024},
doi     = {10.1145/3643764},
url     = {https://doi.org/10.1145/3643764}
}

@misc{snyk-litellm-2026,
  title        = {Poisoned Security Scanner: Backdooring LiteLLM},
  author       = {{Snyk}},
  year         = {2026},
  howpublished = {\url{https://snyk.io/articles/poisoned-security-scanner-backdooring-litellm/}},
  note         = {Accessed: 2026-03-25}
}

@misc{pep740,
  title        = {PEP 740: Index Support for Digital Attestations},
  author       = {Woodruff, William and Tuesca, Facundo and Ingram, Dustin},
  year         = {2024},
  howpublished = {\url{https://peps.python.org/pep-0740/}},
  note         = {Accessed: 2026-03-25}
}

@misc{reproducible-builds-sde,
  title        = {{SOURCE\_DATE\_EPOCH} — Reproducible Builds},
  author       = {{Reproducible Builds Project}},
  year         = {2023},
  howpublished = {\url{https://reproducible-builds.org/docs/source-date-epoch/}},
  note         = {Accessed: 2026-03-28}
}

@misc{ReproducibleBuildsOrg,
  author = {{Reproducible Builds}},
  title = {Reproducible Builds Website},
  url = {https://reproducible-builds.org/}
}

@misc{axiosPostmortem2026,
  title        = {Post Mortem: axios npm supply chain compromise},
  howpublished = {\url{https://github.com/axios/axios/issues/10636}},
  year         = {2026},
  note         = {Accessed 2026-05-20}
}

@misc{snykNodeIpc2026,
  title        = {Malicious node-ipc Versions Published to npm},
  howpublished = {\url{https://snyk.io/blog/malicious-node-ipc-versions-published-npm/}},
  year         = {2026},
  note         = {Accessed 2026-05-20}
}

@misc{snykTanstack2026,
  title        = {TanStack npm Packages Compromised},
  howpublished = {\url{https://snyk.io/blog/tanstack-npm-packages-compromised/}},
  year         = {2026},
  note         = {Accessed 2026-05-20}
}

@misc{snykAntV2026,
  title        = {Mini Shai-Hulud Hits AntV: 300+ Malicious npm Packages Published via Compromised Maintainer Account},
  howpublished = {\url{https://snyk.io/blog/mini-shai-hulud-antv-npm-supply-chain-attack/}},
  year         = {2026},
  note         = {Accessed 2026-05-20}
}

@misc{stepsecurityDurabletask2026,
  title        = {Microsoft's durabletask PyPI Package Compromised in Supply Chain Attack},
  howpublished = {\url{https://www.stepsecurity.io/blog/microsofts-durabletask-pypi-package-compromised-in-supply-chain-attack}},
  year         = {2026},
  note         = {Accessed 2026-05-20}
}

@online{diffoscope,
  title        = {diffoscope: in-depth comparison of files, archives, and directories},
  author       = {{Reproducible Builds}},
  url          = {https://diffoscope.org/},
  note         = {Software tool developed as part of the Reproducible Builds project. Accessed: 2026-06-20}
}

@inproceedings{hassanshahiUnlockingReproducibilityAutomating2025,
	title = {Unlocking Reproducibility: Automating re-Build Process for Open-Source Software},
	issn = {2643-1572},
	url = {https://ieeexplore.ieee.org/abstract/document/11334351},
	doi = {10.1109/ASE63991.2025.00280},
	shorttitle = {Unlocking Reproducibility},
	eventtitle = {2025 40th {IEEE}/{ACM} International Conference on Automated Software Engineering ({ASE})},
	pages = {3392--3402},
	booktitle = {2025 40th {IEEE}/{ACM} International Conference on Automated Software Engineering ({ASE})},
	author = {Hassanshahi, Behnaz and Mai, Trong Nhan and Smith, Benjamin Selwyn and Allen, Nicholas},
	urldate = {2026-06-26},
	date = {2025-11},
	year = {2025},
	note = {{ISSN}: 2643-1572},
}

@inproceedings{decarnedecarnavaletChallengesImplicationsVerifiable2014,
	location = {New York, {NY}, {USA}},
	title = {Challenges and implications of verifiable builds for security-critical open-source software},
	isbn = {978-1-4503-3005-3},
	url = {https://dl.acm.org/doi/10.1145/2664243.2664288},
	doi = {10.1145/2664243.2664288},
	series = {{ACSAC} '14},
	pages = {16--25},
	booktitle = {Proceedings of the 30th Annual Computer Security Applications Conference},
	publisher = {Association for Computing Machinery},
	author = {de Carné de Carnavalet, Xavier and Mannan, Mohammad},
	urldate = {2026-06-26},
	date = {2014-12-08},
	year = {2014},
}

@inproceedings{xiongBuildVerifiabilityJavabased2022,
	location = {New York, {NY}, {USA}},
	title = {Towards build verifiability for Java-based systems},
	isbn = {978-1-4503-9226-6},
	url = {https://dl.acm.org/doi/10.1145/3510457.3513050},
	doi = {10.1145/3510457.3513050},
	series = {{ICSE}-{SEIP} '22},
	pages = {297--306},
	booktitle = {Proceedings of the 44th International Conference on Software Engineering: Software Engineering in Practice},
	publisher = {Association for Computing Machinery},
	author = {Xiong, Jiawen and Shi, Yong and Chen, Boyuan and Cogo, Filipe R. and Jiang, Zhen Ming (Jack)},
	urldate = {2026-06-26},
	date = {2022-10-17},
	year = {2022},
}

@inproceedings{malkaDoesFunctionalPackage2025,
	title = {Does Functional Package Management Enable Reproducible Builds at Scale? Yes.},
	issn = {2574-3864},
	url = {https://ieeexplore.ieee.org/abstract/document/11025777},
	doi = {10.1109/MSR66628.2025.00115},
	shorttitle = {Does Functional Package Management Enable Reproducible Builds at Scale?},
	eventtitle = {2025 {IEEE}/{ACM} 22nd International Conference on Mining Software Repositories ({MSR})},
	pages = {775--787},
	booktitle = {2025 {IEEE}/{ACM} 22nd International Conference on Mining Software Repositories ({MSR})},
	author = {Malka, Julien and Zacchiroli, Stefano and Zimmermann, Théo},
	urldate = {2026-06-26},
	date = {2025-04},
	year = {2025},
	note = {{ISSN}: 2574-3864},
}

@inproceedings{randrianainaOptionsMatterDocumenting2024,
	location = {New York, {NY}, {USA}},
	title = {Options Matter: Documenting and Fixing Non-Reproducible Builds in Highly-Configurable Systems},
	isbn = {979-8-4007-0587-8},
	url = {https://dl.acm.org/doi/10.1145/3643991.3644913},
	doi = {10.1145/3643991.3644913},
	series = {{MSR} '24},
	shorttitle = {Options Matter},
	pages = {654--664},
	booktitle = {Proceedings of the 21st International Conference on Mining Software Repositories},
	publisher = {Association for Computing Machinery},
	author = {Randrianaina, Georges Aaron and Khelladi, Djamel Eddine and Zendra, Olivier and Acher, Mathieu},
	urldate = {2026-06-26},
	date = {2024-07-02},
	year = {2024},
}

@inproceedings{malkaReproducibilityBuildEnvironments2024,
	location = {New York, {NY}, {USA}},
	title = {Reproducibility of Build Environments through Space and Time},
	isbn = {979-8-4007-0500-7},
	url = {https://dl.acm.org/doi/10.1145/3639476.3639767},
	doi = {10.1145/3639476.3639767},
	series = {{ICSE}-{NIER}'24},
	pages = {97--101},
	booktitle = {Proceedings of the 2024 {ACM}/{IEEE} 44th International Conference on Software Engineering: New Ideas and Emerging Results},
	publisher = {Association for Computing Machinery},
	author = {Malka, Julien and Zacchiroli, Stefano and Zimmermann, Théo},
	urldate = {2026-06-26},
	date = {2024-05-24},
	year = {2024},
}

@online{WhyReproducibleBuilds,
	title = {Why reproducible builds? — reproducible-builds.org},
	url = {https://reproducible-builds.org/docs/why/},
	urldate = {2026-06-26},
}

@online{BestpracticesSoftwaresupplychainOssscbestpracticesmd,
	title = {Open Source Software Supply Chain Best Practices at the Eclipse Foundation},
	url = {https://github.com/eclipse-cbi/best-practices/blob/main/software-supply-chain/osssc-best-practices.md},
	titleaddon = {{GitHub}},
	urldate = {2026-06-26},
	langid = {english},
}

@online{CISANSAODNI2022,
	title = {{CISA}, {NSA}, and {ODNI} Release Part One of Guidance on Securing the Software Supply Chain {\textbar} {CISA}},
	url = {https://www.cisa.gov/news-events/alerts/2022/09/02/cisa-nsa-and-odni-release-part-one-guidance-securing-software-supply-chain},
	urldate = {2026-06-26},
	date = {2022-09-02},
	year = {2022},
	langid = {english},
}

@online{sigOpenSourceProject,
	title = {Open Source Project Security Baseline},
	url = {https://baseline.openssf.org/versions/2025-02-25.html},
	titleaddon = {Open Source Project Security Baseline},
	author = {{SIG}, {OpenSSF} Security Baseline},
	urldate = {2026-06-26},
	langid = {english},
}

@misc{moore__software_2024,
  author       = {Marina Moore},
  title        = {Software Supply Chain Security Best Practices v2},
  year         = {2024},
  month        = nov,
  howpublished = {CNCF TAG Security},
  url          = {https://tag-security.cncf.io/blog/software-supply-chain-security-best-practices-v2/},
  note         = {Accessed: 2026-06-26}
}

@article{gao_pyradar_2024,
	title = {{PyRadar}: Towards Automatically Retrieving and Validating Source Code Repository Information for {PyPI} Packages},
	volume = {1},
	url = {https://dl.acm.org/doi/10.1145/3660822},
	doi = {10.1145/3660822},
	shorttitle = {{PyRadar}},
	pages = {115:2608--115:2631},
	issue = {{FSE}},
	journaltitle = {Proceedings of the {ACM} on Software Engineering},
	journal = {Proceedings of the {ACM} on Software Engineering},
	shortjournal = {Proc. {ACM} Softw. Eng.},
	author = {Gao, Kai and Xu, Weiwei and Yang, Wenhao and Zhou, Minghui},
	urldate = {2026-06-26},
	date = {2024-07-12},
	year = {2024},
}

@article{decan_empirical_2019,
	title = {An empirical comparison of dependency network evolution in seven software packaging ecosystems},
	volume = {24},
	issn = {1573-7616},
	url = {https://doi.org/10.1007/s10664-017-9589-y},
	doi = {10.1007/s10664-017-9589-y},
	pages = {381--416},
	number = {1},
	journaltitle = {Empirical Software Engineering},
	journal = {Empirical Software Engineering},
	shortjournal = {Empir Software Eng},
	author = {Decan, Alexandre and Mens, Tom and Grosjean, Philippe},
	urldate = {2026-06-26},
	date = {2019-02-01},
	year = {2019},
	langid = {english},
}

@online{noauthor_debian_nodate,
	title = {Debian -- The universal operating system},
	url = {https://www.debian.org/},
	urldate = {2026-06-26},
}

@online{noauthor_npm_nodate,
	title = {npm {\textbar} Home},
	url = {https://www.npmjs.com/},
	urldate = {2026-06-26},
	langid = {english},
}

@online{noauthor_pypi_nodate,
	title = {{PyPI} · The Python Package Index},
	url = {https://pypi.org/},
	titleaddon = {{PyPI}},
	urldate = {2026-06-26},
	langid = {english},
}

@online{noauthor_trusted_nodate,
	title = {Trusted publishing for npm packages {\textbar} npm Docs},
	url = {https://docs.npmjs.com/trusted-publishers},
	urldate = {2026-06-26},
	langid = {english},
}

@online{noauthor_supply_nodate,
	title = {Supply chain threats},
	url = {https://slsa.dev/spec/v1.0/threats-overview},
	titleaddon = {{SLSA}},
	urldate = {2026-06-26},
	langid = {english},
}

@inproceedings{duan_towards_2021,
	location = {Virtual},
	title = {Towards Measuring Supply Chain Attacks on Package Managers for Interpreted Languages},
	isbn = {978-1-891562-66-2},
	url = {https://www.ndss-symposium.org/wp-content/uploads/ndss2021_1B-1_23055_paper.pdf},
	doi = {10.14722/ndss.2021.23055},
	eventtitle = {Network and Distributed System Security Symposium},
	booktitle = {Proceedings 2021 Network and Distributed System Security Symposium},
	publisher = {Internet Society},
	author = {Duan, Ruian and Alrawi, Omar and Kasturi, Ranjita Pai and Elder, Ryan and Saltaformaggio, Brendan and Lee, Wenke},
	urldate = {2026-06-26},
	date = {2021},
	year = {2021},
	langid = {english},
}

@inproceedings{ohm_backstabbers_2020,
	location = {Cham},
	title = {Backstabber’s Knife Collection: A Review of Open Source Software Supply Chain Attacks},
	isbn = {978-3-030-52683-2},
	doi = {10.1007/978-3-030-52683-2_2},
	url = {https://doi.org/10.1007/978-3-030-52683-2_2},
	shorttitle = {Backstabber’s Knife Collection},
	pages = {23--43},
	booktitle = {Detection of Intrusions and Malware, and Vulnerability Assessment},
	publisher = {Springer International Publishing},
	author = {Ohm, Marc and Plate, Henrik and Sykosch, Arnold and Meier, Michael},
	editor = {Maurice, Clémentine and Bilge, Leyla and Stringhini, Gianluca and Neves, Nuno},
	date = {2020},
	year = {2020},
	langid = {english},
}

@misc{pandyaMaliciousDYdXPackages2026,
  author       = {Kush Pandya},
  title        = {Malicious {dYdX} Packages Published to npm and {PyPI} After Maintainer Compromise},
  year         = {2026},
  month        = feb,
  howpublished = {Socket},
  url          = {https://socket.dev/blog/malicious-dydx-packages-published-to-npm-and-pypi},
  note         = {Accessed: 2026-06-26}
}

@online{noauthor_account_nodate,
	title = {Account Takeover and Malicious Replacement of ctx Project — Python Security 0.0 documentation},
	url = {https://python-security.readthedocs.io/pypi-vuln/index-2022-05-24-ctx-domain-takeover.html},
	urldate = {2026-06-26},
}

@inproceedings{dietrich_levels_2025,
	title = {Levels of Binary Equivalence for the Comparison of Binaries from Alternative Builds},
	issn = {2576-3148},
	url = {https://ieeexplore.ieee.org/abstract/document/11186060},
	doi = {10.1109/ICSME64153.2025.00058},
	eventtitle = {2025 {IEEE} International Conference on Software Maintenance and Evolution ({ICSME})},
	pages = {576--587},
	booktitle = {2025 {IEEE} International Conference on Software Maintenance and Evolution ({ICSME})},
	author = {Dietrich, Jens and White, Tim and Hassanshahi, Behnaz and Krishnan, Paddy},
	urldate = {2026-06-30},
	date = {2025-09},
	year = {2025},
	note = {{ISSN}: 2576-3148},
}

@misc{DebianReproducibleTests,
  author       = {{Debian Reproducible Builds Team}},
  title        = {Debian Variance Testing Dashboard},
  year         = {2026},
  url          = {https://tests.reproducible-builds.org/debian/reproducible.html},
  note         = {Accessed: 2026-06-30}
}

@online{noauthor_securing_nodate,
	title = {Securing Build Pipelines},
	url = {https://tag-security.cncf.io/community/publications/supply-chain-security-tools/securing-build-pipelines/},
	titleaddon = {{CNCF} {TAG} Security},
	urldate = {2026-06-30},
	langid = {english},
	note = {Section: community},
}

@online{noauthor_variations_nodate,
	title = {Variations in the build environment — reproducible-builds.org},
	url = {https://reproducible-builds.org/docs/env-variations},
	urldate = {2026-06-30},
}

@online{noauthor_build_nodate,
	title = {Build: Verifying artifacts},
	url = {https://slsa.dev/spec/v1.2/verifying-artifacts},
	shorttitle = {Build},
	titleaddon = {{SLSA}},
	urldate = {2026-06-30},
	langid = {english},
}

@online{noauthor_introducing_2025,
	title = {Introducing {OSS} Rebuild: Open Source, Rebuilt to Last},
	url = {https://blog.google/security/introducing-oss-rebuild-open-source/},
	shorttitle = {Introducing {OSS} Rebuild},
	titleaddon = {Google},
	urldate = {2026-07-01},
	date = {2025-07-21},
	year = {2025},
	langid = {english},
}

@online{openssf_introducing_nodate,
	title = {Introducing Artifact Attestations—Now in Public Beta – Open Source Security Foundation},
	url = {https://openssf.org/blog/2024/05/24/introducing-artifact-attestations-now-in-public-beta/},
	author = {{OpenSSF}},
	urldate = {2026-06-30},
	langid = {american},
}

@online{noauthor_provenance_nodate,
	title = {Provenance},
	url = {https://slsa.dev/spec/v1.0/provenance},
	titleaddon = {{SLSA}},
	urldate = {2026-07-01},
	langid = {english},
}

@software{nektos_act,
  author       = {{Nektos Authors}},
  title        = {act: Run your GitHub Actions locally},
  month        = jul,
  year         = {2026},
  publisher    = {GitHub},
  version      = {0.2.70},
  url          = {https://github.com/nektos/act}
}

@inproceedings{okafor_sok_2022,
  author    = {Chinenye Okafor and Taylor R. Schorlemmer and
               Santiago Torres-Arias and James C. Davis},
  title     = {{SoK}: Analysis of Software Supply Chain Security by
               Establishing Secure Design Properties},
  booktitle = {Proceedings of the 2022 ACM Workshop on Software Supply
               Chain Offensive Research and Ecosystem Defenses},
  series    = {SCORED '22},
  pages     = {15--24},
  year      = {2022},
  publisher = {Association for Computing Machinery},
  doi       = {10.1145/3560835.3564556}
}

@misc{sonatype_state_2026,
  author       = {{Sonatype}},
  title        = {2026 State of the Software Supply Chain},
  year         = {2026},
  howpublished = {Sonatype},
  url          = {https://www.sonatype.com/state-of-the-software-supply-chain/introduction},
  note         = {Accessed: 2026-08-11}
}
\clearpage

\end{document}